\documentclass[12pt]{article}
\usepackage{axodraw,epsfig}
\usepackage{graphics}

\catcode`@=11
\newcount\@tempcntc
\def\@citex[#1]#2{\if@filesw\immediate\write\@auxout{\string\citation{#2}}\fi
  \@tempcnta\z@\@tempcntb\m@ne\def\@citea{}\@cite{\@for\@citeb:=#2\do
    {\@ifundefined
       {b@\@citeb}{\@citeo\@tempcntb\m@ne\@citea\def\@citea{,}{\bf ?}\@warning
       {Citation `\@citeb' on page \thepage \space undefined}}%
    {\setbox\z@\hbox{\global\@tempcntc0\csname b@\@citeb\endcsname\relax}%
     \ifnum\@tempcntc=\z@ \@citeo\@tempcntb\m@ne
       \@citea\def\@citea{,}\hbox{\csname b@\@citeb\endcsname}%
     \else
      \advance\@tempcntb\@ne
      \ifnum\@tempcntb=\@tempcntc
      \else\advance\@tempcntb\m@ne\@citeo
      \@tempcnta\@tempcntc\@tempcntb\@tempcntc\fi\fi}}\@citeo}{#1}}
\def\@citeo{\ifnum\@tempcnta>\@tempcntb\else\@citea\def\@citea{,}%
  \ifnum\@tempcnta=\@tempcntb\the\@tempcnta\else
   {\advance\@tempcnta\@ne\ifnum\@tempcnta=\@tempcntb \else \def\@citea{--}\fi
    \advance\@tempcnta\m@ne\the\@tempcnta\@citea\the\@tempcntb}\fi\fi}
\catcode`@=12

\begin{document}

\begin{titlepage}

    \begin{flushright}
      \normalsize LPT-ORSAY/08-19 
      %hep-ph/yymmnnn [to be added later]\\
      %18 June 2007  
    \end{flushright}

\vskip1.5cm
\begin{center}
\Large\bf\boldmath
Parametrisations of the $D\to K\ell\nu$ form factor\\
and the determination of $\hat{g}$
\unboldmath
\end{center}

\vspace*{0.8cm}
\begin{center}

{\sc S\'ebastien Descotes-Genon} and {\sc Alain Le Yaouanc}\\[5mm]
  {\small\it Laboratoire de Physique 
Th\'eorique,\\[0.1cm]
CNRS/Univ. Paris-Sud 11 (UMR 8627),
91405 Orsay Cedex, France}

\end{center}

\vspace*{0.8cm}
\begin{abstract}
  \noindent
The vector form factor $f_+(t)$ of the semileptonic decay $D\to K\ell\nu$, 
measured recently with a high accuracy, can be used to determine 
the strong coupling constant $g_{D_s^* D K}$. The latter is related to
the normalised coupling $\hat g$ releveant in heavy-meson chiral perturbation
theory. This determination relies on the estimation
of the residue of the form factor at
the $D_s^*$ pole and thus on an extrapolation of the form factor
in the unphysical region $(m_D-m_K)^2<t<(m_D+m_K)^2$. 
We test this extrapolation for several parametrisations of the form
factors by determining the value of $\hat{g}$, whose value can be compared to
other (experimental and theoretical) estimates.
Several unsophisticated parametrisations, differing by the amount of physical
information that they embed, are shown to pass this test.
An apparently more elaborated parametrisation 
of form factors, the so-called $z$-expansion, is at variance with the other
models, and we point out some significant shortcomings of this parametrisation
for the problem under consideration.

%\vspace*{0.8cm}
%\noindent
%PACS numbers:

\end{abstract}

\vfil
\end{titlepage}

\newpage

The weak transitions from one meson to another provide very interesting tests
of our understanding of the Standard Model both in the weak and strong
sectors. On one hand, elements of the Cabibbo-Kobayashi-Maskawa matrix
can be determined by comparing the experimentally measured decay rate with 
a theoretical value obtained at one or several kinematic points: for instance
$B\to D\ell\nu$ for $|V_{cb}|$, $B\to \pi\ell\nu$ for $|V_{ub}|$, or
$K\to\pi\ell\nu$ for $|V_{us}|$. On the other hand, the shape of the spectrum
provides a stringent check of our description of the dynamics of hadrons
governed by QCD. 

The hadronic physics of the problem is encoded in form
factors, which are complicated nonperturbative objects. 
Due to our limited theoretical knowledge of these objects, many
parametrisations are available, trying to include as much information
as possible on the (known or assumed) dynamics of the corresponding mesons. 
Recently, the BaBar collaboration has 
performed a very accurate analysis of the $D^0\to K^- e^+\nu_e$ 
decay~\cite{Aubert:2007wg} (see also 
the work done by the CLEO collaboration~\cite{CLEO:2007sm}). 
The relevant form factors are defined as
\begin{equation}
\langle K(p')|\bar{s}\gamma_\mu c| D(p)\rangle
 = \left(p_\mu + p'_\mu - q_\mu \frac{m_D^2-m_K^2}{q^2}\right)f_+(t)
    + q_\mu \frac{m_D^2-m_K^2}{q^2} f_0(t)
\end{equation}
with $q=p-p'$ and $t=q^2$.
Ref.~\cite{Aubert:2007wg} determined very precisely the 
$t$-dependence of the vector form factor $f_+$ (the scalar form factor $f_0$
could not be obtained because of the very light mass of the electron).
We can provide a general expression of this form factor in terms of an unsubtracted
dispersion relation (such an unsubtracted dispersion relation is allowed by
the asymptotic behaviour expected from QCD): 
\begin{equation} \label{RD}
f_+(t)=\frac{{\rm Res}(f_+)}{m_{D_s^*}^2-t}
       +\frac{1}{\pi}\int_{t_+}^\infty dt' \frac{{\rm Im}\ f_+(t')}{t'-t}
\end{equation}
where
\begin{equation}
t_+ = (m_D+m_K)^2 \qquad t_-=(m_D-m_K)^2
\end{equation}
meaning that we have a pole at $m_{D_s^*}^2$ ($m_{D_s^*}=2.112$~GeV) and a cut from
the $DK$ continuum~\footnote{In principle, the cut could begin 
at a much lower value, i.e. $D_s\pi^0$, near the $D_s^*$. Indeed, 
the $D_s\pi^0$ state has the appropriate quantum numbers to contribute
to the form factor. However, $D_s\pi^0\to DK$ scattering is exotic and there
is no resonance able to yield a sizeable amplitude for this process. 
Therefore, this cut is likely tiny below the $DK$ threshold, and we will not
consider it in the following.}. 
We have defined ${\rm Res}(f_+)$ so that it be positive. 

The physical region for the semileptonic decay is $0<t<t_-$,
where high-precision experimental information is available~\cite{Aubert:2007wg}.
Although it corresponds to an unphysical point between
the semileptonic region and the $DK$ cut, the pole residue is physically
interesting. Indeed it is related to the  $D$-$K$-$D^*_s$ strong coupling 
(with a subsequent weak decay of $D^*_s$ into a lepton pair), which itself 
linked to $g_{D^* D\pi}$ by $SU(3)$ symmetry. 
This latter coupling is a fundamental
ingredient for Heavy Meson Chiral Perturbation 
Theory~\cite{Wise:1992hn,Yan:1992gz,Burdman:1992gh,Casalbuoni:1996pg} 
and allows one to compute 
the effect of pion and kaon exchanges on the non-perturbative dynamics
of heavy-light mesons (see refs.~\cite{Fajfer:2004mv,Becirevic:2006me,Becirevic:2007dg} 
for recent applications to semileptonic decays and other processes, 
as well as pending issues in this field).

We want to exploit the good experimental knowledge on $f_+$ in the physical
region to determine
the residue ${\rm Res}(f_+)$ and thus $g_{D^*_s DK}$ by an extrapolation of 
the form factor from $t_-$ to $m_{D_s^*}^2$:
\begin{equation} \label{eq:poleg}
f_+(t)=\frac{f_{D_s^*} g_{D^*_s DK}/2m_{D_s^*}}{1-t/m_{D_s^*}^2}+\ldots
\end{equation}
where $f_{D_s^*}$ is the decay constant relative to the purely leptonic decay of the $D_s^*$.
$g_{D^*_s DK}$ is related to another significant quantity, the normalised
matrix element of the axial current between the $B$ and $B^{*}$ mesons denoted
$\hat{g}$.

For this extrapolation, we can devise analytical models or parametrisations
describing the physical region as a function of $t$, and extrapolate the expression into the region near the pole. We then deduce the residue according to:  
\begin{equation}
{\rm Res}(f_+)=\lim_{t \to m_{D_s^*}^2}(m_{D_s^*}^2-t)f_+(t) 
\end{equation}
It proves interesting to define a function ${\rm Res}(f_+)(t)$, which we call the \emph{residue function}, 
according to:
\begin{equation}
{\rm Res}(f_+)(t)=(m_{D_s^*}^2-t)f_+(t) \qquad \qquad {\rm Res}(f_+)={\rm Res}(f_+)(m_{D_s^*}^2)
\end{equation}
We can hope to get reasonable extrapolations by considering \emph{smooth
representations of ${\rm Res}(f_+)(t)$}. Since we have factorised the pole
denominator, the residue function has singularities only due to the cut.

Models will show differences in the way that they represent these
singularities. 
We propose to use the value of $\hat{g}$ as a \emph{test} to select the most
appropriate models among the available ones.
Indeed, we have independent experimental and theoretetical information
on this coupling. Models with reasonable physical assumptions (i.e. the
singularities along the cut) should be able to provide reasonable
extrapolations from the semileptonic region to the $D_s^*$ pole, and thus
to yield values of $\hat{g}$ in good agreement with our current knowledge.

Of course, the various heavy-to-light semileptonic decays could yield similar
strong couplings, somewhat related by heavy quark or $SU(3)$ symmetries. To
evaluate the specific interest of the reaction $D \to K \ell \nu$ in this
respect, we can notice the following facts. Since
\begin{equation}
t_-/m_{D_s^*}^2=0.42 \qquad
  t_+/m_{D_s^*}^2=1.24
\end{equation}
the $D_s^*$ pole is relatively close to the threshold of the cut for
$DK$ production, but far from the physical region for the semileptonic decay;
in addition, the physical interval is not very large. This can be compared
with other semileptonic heavy-to-light decays.
$B,D \to \pi \ell \nu$, exhibit the advantage of a pole located
very close to the physical region, and specifically for $B \to \pi \ell \nu$,
the physical region is very extended. However the cut is now so close to the
investigated pole that it may cause problems. For $B \to \pi l \nu$, the 
smallness of $|V_{ub}|$ makes an accurate measurement of the
form factor near the $t=t_-$ endpoint very difficult: the decay rate is very
small at this endpoint where its value is crucial for a good extrapolation. 

$D \to \pi l \nu$ seems more promising in this respect,
and it is complemented interestingly by $D\to K \ell \nu$. In the latter
case, the pole lies farther than for
$D \to \pi l \nu$, but it is better separated from the cut, and it is easier
to observe because of the large value of the CKM matrix angle $V_{cs}$
(compared to the Cabibbo-suppressed $D\to\pi$ transition).
On one hand this could be considered
unfavourable: the extrapolation will remain undoubtedly uncertain, given the
limited precision of the data, since many models may be very close
in the physical region, but will greatly differ near the pole. On the other
hand, it provides an interesting opportunity to test these parametrisations in
a situation where their differences will be enhanced and thus easier to discuss.

This note is organised as follows. In Sec.~1, we discuss the essential
ingredient of the dispersive representation of the form factor, namely 
the $DK$ cut. In Sec.~2, we introduce several parametrisations which differ
mainly through their approximate description of the cut. In Sec.~3, we discuss
the extrapolation of the form factor in the unphysical region according to
these parametrisations. In Sec.~4, we collect the resulting 
values for the hadronic coupling constants $g_{D_s^* D K}$ and $\hat{g}$
and comment on the discrepancies induced by the different parametrisations,
before drawing a few conclusions in Sec.~5.

\section{The $DK$ cut}

Theoretical inputs concerning the cut associated with $DK$ production
are essential since it governs the behaviour of the residue function ${\rm Res}(f_+)$.
There are none compelling. Yet we can consider two partial and complementary
contributions from the low-energy $DK$ continuum and from resonances.

First, consider the contribution to the cut stemming from 
the $DK$ low-energy continuum. 
In the elastic regime, between the $DK$ threshold and the first inelastic 
threshold, one has
\begin{equation}
{\rm Im} f_+(t)=\sqrt{\lambda}{t}f_1^{DK}(t) f_+(t)\theta(t-(M_D+M_K)^2)
\end{equation}
where $\lambda=(t-t_+)(t-t_-)=4 t q^2$ and
$f_1^{DK}$ is the $P$-wave amplitude for $DK\to DK$ scattering. 
Close to the $DK$ threshold (small positive $q^2$), the form factor has the following behaviour
in $q^2$
\begin{equation}
f_1^{DK}(t)\sim A_1 q^2
\end{equation}
so that
\begin{equation}
{\rm Im\ }f_+^{DK}(t) \sim A_1 q^3
\end{equation}
%and then for $t<t_+$:
%\begin{equation}
%***
%{\rm Re\ }f_+^{DK}(t) \sim A_1 (t_+-t)^{3/2} 
%\end{equation}
which means that the imaginary part departs slowly from zero above the
$DK$ threshold: the cut should be very smooth.

Secondly, there are resonances all along the cut, corresponding to the
successive radial excitations of the $D_s^*$, the first of which should 
be close to the $DK$ threshold. Indeed, with an excitation energy of order
$0.5$~GeV added to the $D_s^*$ mass~\footnote{This order of magnitude for the
  excitation energy is found in most quark models, see for instance
  ref.~\cite{Becirevic:1999fr} discussed below.}, we get a first radial
excitation around $2.6$~GeV, while the cut starts at $m_D+m_K \simeq
2.4$~GeV. If we worked in the large-$N_c$ approximation of resonances with 
vanishing widths, the cut of the form factor $f_+$ would be largely
dominated by the pole of the first excitation. On the other hand, in the
actual world, most of the resonances acquire a broad width through their 
strong decays, will overlap and interfere, yielding a rather smooth
cut. This is presumably true even for the lowest radial
excitation, if the coupling constant is not exceptionally small, because of the allowed phase space, so that the impact of this first resonance on the cut is
presumably mild. These arguments are supported by the recent observation
of the $D_{sJ}(2700)^+$ by the Belle collaboration~\cite{:2007aa}. This meson
is interpreted as a radially excited $c\bar{s}$ state with $J^P=1^-$ with 
a mass $M=2708\pm 9^{+11}_{-10}$ MeV, quite far away from the start of the 
cut, and a fairly broad width $\Gamma=108\pm 23^{+36}_{31}$ MeV.

Even though we have arguments supporting the smooth rise of the cut above
threshold, 
it would be useful to estimate the \emph{absolute strength} of the cut in
order to constrain our extrapolation more tightly. Ideally, a determination
of the vacuum-to-$DK$ matrix element would provide the basis for an interpolation
between the physical region for the semileptonic and the region
of the cut, rather than an extrapolation. Since this piece of information is
currently lacking, we have to rely on parametrisations of the form factor.

\section{Parametrisations of the form factor}

We speak of ``parametrisations'' because we have admittedly little theory
behind them. They are functions of $t$ with the following main merits:  1) in
the physical region  they collect the data of the form factors in a both
simple and accurate description; 2) they incorporate in a rough way the few
safe statements that we can formulate about the form factors outside the
physical region, so that they may still be trusted at least at a certain
distance from the physical region (of course, it remains an assumption that we
can trust them as far as the pole). As
summarised in eq.~(\ref{RD}), we know that there is a pole at the known 
value $t=m_{D_s^*}^2$, and a cut begins at $t=(m_{D_s^*}+m_K)^2$. The various
parametrisations propose explicitly or implicitly different simplifications in the treatment of the
cut, and we will sketch the salient features of the most
widely used ones. 

In the case at hand, $D \to K \ell \nu$, the experimental data is so
accurate that the best option consists in fitting the parameters 
of the parametrisations on the data in the physical region. It turns out that 
most of them fit the data equally well in the semileptonic region. 
A particular representation would exhibit
a decisive advantage if it were able to describe the unphysical region between
the semileptonic region and the cut \emph{with the same parameters}.

\subsection{$z$-expansion}

A parametrisation which is by now rather popular, yet apparently
sophisticated, is the $z$-expansion~\cite{Hill:2006ub,Hill:2006bq},
which dates back to Boyd et al. (see ref.~\cite{Boyd:1997qw} and the works with
Grinstein and Lebed quoted therein) where it was applied to heavy-to-heavy
transitions. In  principle, it has been devised to
offer a good representation of form factors and we recall some steps of its
derivation below. 

One aims at writing down $f_+$ as a series $\Sigma~a_i
z(t)^i$. The expansion parameter $z(t)$ is defined as:
\begin{equation}
z(t,t_0)=\frac{\sqrt{t_+-t}-\sqrt{t_+-t_0}}{\sqrt{t_+-t}+\sqrt{t_+-t_0}}
       \propto t_0-t
\end{equation}
with a free parameter $t_0$. $z(t)$
remains small over the whole physical region, and maps 
the complex $t$-plane into a
disk of radius 1, with the cut in $t$ being transformed 
into the circle $|z|=1$.
Ref.~\cite{Hill:2006ub} advocates the choice $t_0/t_+=1-\sqrt{1-t_-/t_+}$ 
($t_0/m_{D_s^*}^2=0.23$), which implies that $z(t)=0$ occurs
for $t$ in the middle of the physical region [$z(0)=0.0515$ and
$z(t_-)=-0.0515$]. The change of variables from $t$ to $z$ has often been used
in the hope of getting a quicker convergence of the resulting series 
in the physical region.
  
It has been proposed to combine this idea with
analyticity constraints in order to constrain further
the coefficients of the $z$-expansion~\cite{Hill:2006ub,Hill:2006bq,Boyd:1997qw}. The
starting point is the correlator
\begin{equation}
\Pi^{\mu\nu}(q)=i\int d^4x\ e^{iqx} 
 \langle 0|T(\bar{s}\gamma^\mu c)(x)\ (\bar{s}\gamma^\nu c)^\dag(0) 
        |0 \rangle
 =(q^\mu q^\nu - q^2 g^{\mu\nu}) \Pi^T(q^2) + g^{\mu\nu} \Pi^L(q^2)
\end{equation}
which defines two polarisation functions (transverse and longitudinal).
First, these functions can be bounded approximately at large momenta 
using the Operator Product Expansion. Second,
these functions 
can be expressed in terms of their imaginary part through a dispersion
relation. Using unitarity, one can express 
this imaginary part as a sum of various (positive) contributions,
corresponding to the squared modulus of matrix elements 
$\langle H |\bar{s}\gamma^\mu c|0\rangle$, where $H$ denotes any
arbitrary single- or multiple-particle state 
with the appropriate quantum numbers (each contribution contains a
a factor coming from the phase space). 
It means in particular that the imaginary part is larger than the sole 
contribution from $H=DK$, i.e. from $f_+$, multiplied
by a phase-space factor. 

By choosing multiplying the series in $z$ by a carefully chosen factor
denoted $\Phi$, one can convert the bound induced on $f_+$ by unitarity and
OPE into a bound on the coefficients of the series in $z$. 
The $z$-representation can be written as follows:  
\begin{equation}
z{\rm{-exp}}: \frac{f_+(t)}{f_+(0)} = 
  \frac{P(0)\Phi(0,t_0)}{P(t)\Phi(t,t_0)} \times 
        \frac{1+\frac{a_1}{a_0} z(t,t_0)+\frac{a_2}{a_0} z^2(t,t_0)}
             {1+\frac{a_1}{a_0} z(0,t_0)+\frac{a_2}{a_0} z^2(0,t_0)}
\end{equation}
where  
\begin{equation} \label{Phi}
\Phi(t,t_0) = N (t_+-t) (\sqrt{t_+ -t}+\sqrt{t_+})^{-5}
                (\sqrt{t_+ -t}+\sqrt{t_+-t_0})
                (\sqrt{t_+ -t}+\sqrt{t_+-t_-})^{3/2}
\end{equation}
where $N$ is a numerical normalisation factor. It is unimportant here, 
since we consider only ratios of the form factor to its value at $q^2=0$.
The factor $P(t)=z(t,m_{D_s^*}^2)$ includes the $D_s^*$ pole.
As explained above, the coefficients $a_n$ are bounded by unitarity:
$\sum~|a_i|^2<1$ with a properly chosen normalisation constant $N$~\cite{Hill:2006ub,Hill:2006bq,Boyd:1997qw}. 

In heavy-to-light processes, this expansion seems less useful, 
because it appears that all the coefficients $a_n$ that can be
determined are anyway much smaller 
than the unitarity bound in absolute value~\cite{Aubert:2007wg},
and nothing safe can be said about the \emph{relative} magnitude of the
coefficients with respect to the first one (see the fits in sec.~3). 
There remains a useful bound on
the \emph{rest} of the series, obtained by using Cauchy's inequality:
\begin{equation} \label{borne}
\left|\sum_n^{\infty}a_i z^i\right| \leq  
  \sqrt{\sum_n^{\infty}|a_i|^2}\sqrt{\sum_n^{\infty}|z^i|^2} 
     \leq \sqrt{\frac{|z|^{2 n}} {1-|z|^2}}
\end{equation}
This bound, given the observed values of the form factor,
is sufficiently small to be meaningful: it shows that the necessary number of
terms should be about 2 or 3 to get a one percent accuracy in the semileptonic
region. Actually, it was found that 2 or 3 terms are sufficient to fit the data~\cite{Aubert:2007wg},
the last coefficient ($a_2$) being already affected by a large
uncertainty. Unfortunately, this accurate representation in the 
semileptonic region does not warrant an accurate extrapolation. Indeed,
in the region $t_-<t<t_+$, $z(t$) by no means remains small in this region
$t_-<t<t_+$, since at the pole, $z (m_{D_s^*}^2)=-0.340$ and at the beginning of the cut,
$z (t_+)=-1$. In this region, the bound eq.~(\ref{borne}) 
becomes ineffective, or indicates that very many terms could be necessary.

\subsection{Alternative representations}

This parametrisation can be compared with other less sophisticated parametrisations
\begin{eqnarray}
{\rm Pole}: \frac{f_+(t)}{f_+(0)}&=&
\frac{1}{\left[1-\frac{t}{m_{pole}^2}\right]}\\
{\rm BK}: \frac{f_+(t)}{f_+(0)}&=&
\frac{1}{\left[1-\frac{t}{m_{D_s^*}^2}\right]
         \left[1-\alpha\cdot \frac{t}{m_{D_s^*}^2}\right]}\\
{\rm Linear}: \frac{f_+(t)}{f_+(0)}&=&
\frac{1}{1-\frac{t}{m_{D_s^*}^2}}
    \left[1+c_1 \cdot \frac{t}{m_{D_s^*}^2}\right]\\
{\rm Quadratic}: \frac{f_+(t)}{f_+(0)}&=&
\frac{1}{1-\frac{t}{m_{D_s^*}^2}}
  \left[1+c_1 \cdot \frac{t}{m_{D_s^*}^2}+c_2 \cdot \frac{t^2}{m_{D_s^*}^4}\right]
\end{eqnarray}

\begin{itemize}
\item The pole parametrisation is certainly too naive, since there is no
  reason for the lowest lying pole to saturate the form factor.
 Not surprisingly, one finds the pole much below its actual position (the
 $D_s^*$ mass). We will not use this parametrisation to perform an extrapolation 
of $f_+$ to the $D_s^*$ mass; it serves only to show that the 
achieved experimental accuracy requires to go beyond the common
assumption of the dominance by the lowest pole \emph{in the physical region}.

\item The BK parametrisation~\cite{Becirevic:1999kt} is well known in the
  analysis of $B$-decays, and it is mainly motived by the
  scaling laws of the form factors in the heavy quark 
  limit. For a $D$ decay, 
  this motivation is certainly weaker, but it is useful by providing
  a crude representation of the cut by an additional pole, with a free mass
  which comes out neatly \emph{above} the start of the cut, 
  at $t=m_{D_s^*}/\alpha$, with $\alpha$ smaller than $1$ to be
  consistent with the location of the cut. 
Let us stress that this second pole is
not to be identified with any particular resonance. This effective
pole sums up the effect of infinitely many resonances
with positive or negative contribution (note that the additional BK pole
has a contribution opposite in sign to $D_s^*$).

\item Almost equivalent in practice, the "linear" and "quadratic"
  parametrisations follow 
  old ideas from current algebra. They assume that the
amplitude have a polynomial dependence of low degree in the momenta, once the
lowest lying (ground state) poles have been factored out. Such an expansion
is justified in the case of a weak cut, which can be approximated by its
expansion in $t$ with a reasonable accuracy.

\item Another parametrisation can be obtained by considering the $z$-expansion
and replacing $\Phi$ with a constant. Obviously, no unitarity bounds can be
derived in such a case: we have just performed a change of variable
(reexpressing a $t$-series into a $z$-series) and extracted the $D_s^*$ pole. In order to distinguish these
two versions of the $z$-expansion method, we call ``simple $z$-expansion'' the
parametrisation with $\Phi(t)$ set to 1, and ``unitary $z$-expansion'' the
expansion relying on eq.~(\ref{Phi}), which allows to exploit unitarity
constraints (at least in principle) and described for instance in refs.~\cite{Hill:2006ub,Hill:2006bq,Boyd:1997qw}.
\end{itemize}

\section{Fit of the parametrisations to $D\to K\ell\nu$ data}

Before extrapolating these various parametrisation to the $D_s^*$ pole, we
need to determine their parameters from the data in the physical region. 
A fit of $z$-expansion, BK and pole parametrisations has been made
in ref.~\cite{Aubert:2007wg} with data corrected for 
radiative effects. Unfortunately, the data (values of the form factors
and correlation matrix) provided in this reference are only given before 
the correction of radiative effects. According to ref.~\cite{Aubert:2007wg},
the main correction corresponds to an increase of the value of the first bin.
Therefore, for each parametrisation
we can compare three different fits depending on the set of data:
\begin{itemize}
\item The data before radiative corrections and the correlation matrix before
radiative corrections (performed by us),
\item The data with the first bin increased and the correlation matrix before
radiative corrections (performed by us),
\item The data and the correlation matrix with full radiative corrections
(performed in ref.~\cite{Aubert:2007wg} and by us).
\end{itemize}

\begin{table}
\begin{center}
\begin{tabular}{cccccc}
Model & Model param 
          & No radiative corr & First bin corrected & 
               Full radiative corr\\
\hline
Unitary $z$-exp 
   & $a_1/a_0$ & $-2.43 \pm 0.25$ & $-2.43\pm 0.25$ & $-2.5\pm 0.28$\\
   & $a_2/a_0$ & $-4.26 \pm 6.67$ & $-2.98\pm 6.67$ & $0.6\pm 7.8$\\
   & correl    & -0.82 & -0.82 & -0.86\\
\hline
BK & $\alpha$  & $0.38\pm 0.03$ & $0.37\pm 0.04$ & $0.377\pm 0.037$\\
\hline
Pole 
   & $m_{pole}$ & $1.89\pm 0.02$ & $1.89\pm 0.02$ & $1.884\pm 0.019$\\
\hline
Linear & $c_1$ & $0.43\pm 0.04$ & $0.43\pm 0.05$ & $0.42\pm 0.05$\\
\hline
Quadratic & $c_1$ & $0.45\pm 0.15$ & $0.41\pm 0.15$ & $0.33\pm 0.15$\\
          & $c_2$ & $-0.036\pm 0.42$ & $0.046\pm 0.42$ & $0.25 \pm 0.42$\\
          & correl & -0.96 & -0.96 & -0.96\\
\hline
Simple $z$-exp
   & $a_1/a_0$ & $-4.27 \pm 0.26$ & $-4.28\pm 0.26$ & $-4.34\pm 0.26$\\
   & $a_2/a_0$ & $ 3.13 \pm 6.73$ & $ 4.41\pm 6.74$ & $7.78\pm 6.79$\\
   & correl    & -0.85 & -0.85 & -0.85\\
\end{tabular}
\end{center}
\caption{Parameters of the different parametrisations of $f_+$ from 
a fit to $D\to K\ell\nu$ data, with various treatments of radiative
corrections (not included, partially included, fully included).}
\label{tab:fitparam}
\end{table}

The parameters of the different models are collected in
Table~\ref{tab:fitparam}.
Apart from the pole model, which does not fit the data very well,
the other models yield very similar fits.
The corresponding minimal values of the $\chi^2$ remains between 7 and 8,
for 7 or 8 degrees of freedom (fit with 2 or 1 parameter).

The first three entries of the last column are 
taken from ref.~\cite{Aubert:2007wg,Roudeau:2007}, where
we combined systematic and statistic uncertainties in
quadrature. Unfortunately, the three other parametrisations (linear,
quadratic and simple $z$-expansion) have not been considered
in~\cite{Aubert:2007wg}. 
As a poor man's way of getting some information on the impact of radiative corrections for
these two parametrisations, 
we have performed a fit with:
\begin{itemize}
\item The correlation matrix provided in ref.~\cite{Aubert:2007wg}
\item The central values in each bin obtained from the $z$-parametrisation,
using the values of $a_1/a_0$ and $a_2/a_0$ obtained including the
radiative corrections.
\end{itemize}
We have checked that this procedure leads to values and uncertainties for the fitted parameters of
the different models which are very similar to those quoted in
ref.~\cite{Aubert:2007wg}.
However, this procedure is admittedly a very imperfect attempt of getting
a handle on radiative corrections, and should be replaced
by a full treatment of radiative corrections for these three parametrisations,
which can be done only by our experimental colleagues.

\begin{figure}[t]
\begin{center}
\includegraphics[width=12cm]{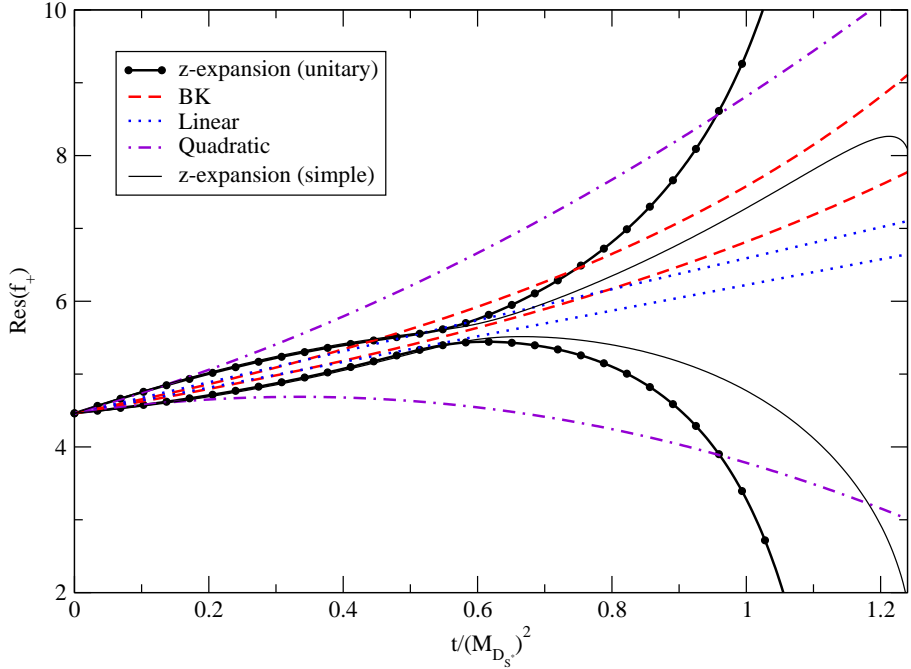}
\caption{Range of variation for the 
function ${\rm Res\ }f_+(t)=f_+(t)(m_{D_s^*}^2-t)$
related to the residue of the form factor at the $D_s^*$ pole as a function
of $t/M_{D_s^*}^2$,
for the different parametrisations considered.
No radiative corrections are included.}
\label{fig:nocorr}
\end{center}
\end{figure}

\begin{figure}[t]
\begin{center}
\includegraphics[width=12cm]{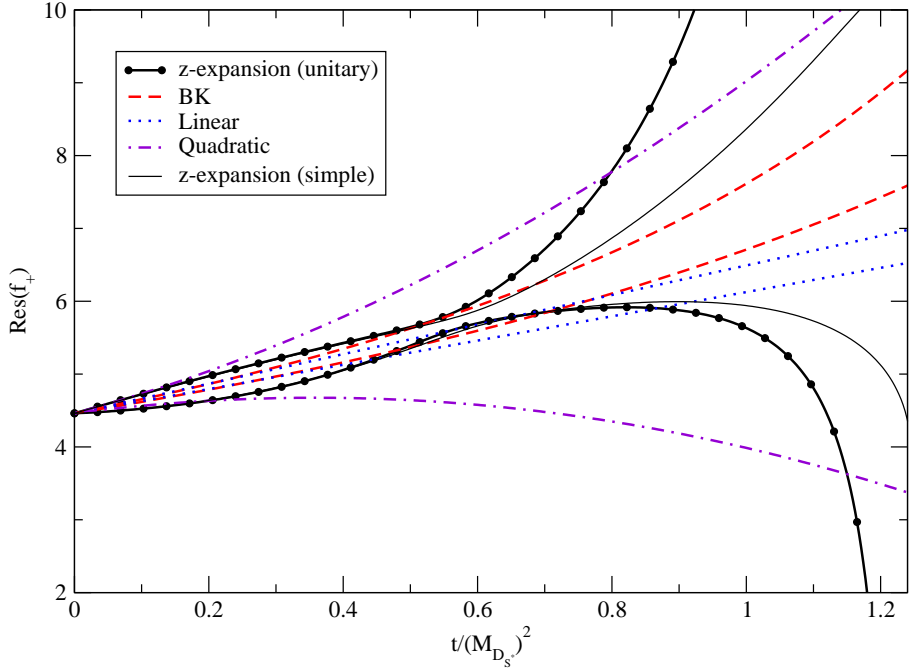}
\caption{Range of variation of the 
function ${\rm Res\ }f_+(t)=f_+(t)(m_{D_s^*}^2-t)$
related to the residue of the form factor at the $D_s^*$ pole, as a function
of $t/M_{D_s^*}^2$,
for the different parametrisations considered.
The effect of radiative corrections is included
according to  ref.~\cite{Aubert:2007wg} in the case
of unitary $z$-expansion and BK models (see the text for
the treatment of radiative corrections for the other models).}
\label{fig:fullcorr}
\end{center}
\end{figure}

Figs.~\ref{fig:nocorr} and \ref{fig:fullcorr} illustrate the range of
variation allowed for the residue function according to the different
parametrisations,  
respectively before and after radiative corrections are taken into account. 
A striking difference can be observed between the (unitary) $z$ representation and
the other parametrisations, the errors in the first case exploding
beyond $t/m_{D_s^*}^2 \simeq 0.8$. 
This behaviour
can be understood from the explicit expression of the $z$-expansion.
One must note the important factor $t_+-t$ in the definition of $\Phi(t,t_0)$,
eq.~(\ref{Phi}). 
This factor implies that the  $z$-expansion includes \emph{a spurious,
  unwanted pole at the threshold of the cut}, a feature which is very
disadvantageous when the expansion is used not far from the cut, as needed to
make our extrapolation. To be used there, the $z$-expansion 
requires in principle a very
large number of terms in the series $\Sigma~a_n z^n$, and of course so many
$a$'s 
cannot be determined from the knowledge of the physical
region. At most, the first two or three coefficients can be estimated, so that
the spurious threshold pole has a very strong effect on extrapolation. 
At the start of the cut, the $z$-expansion becomes meaningless.

The complicated
form of $\Phi$ was chosen to translate the unitarity conditions on $f_+$ into 
a constraint on the coefficients $a_n$ of the $z$-expansion. In particular,
the troublesome threshold pole is related to the two-body phase space arising
in the unitarity bound on $f_+$. On the other hand, it turns out that the bound on the
coefficients $a_n$ is far from saturated, and thus of little impact on the
problem at hand, so that it can be left out without consequence, as
exemplified by the ``simple'' $z$-expansion (with $\Phi=1$).

It is also interesting to compare this parametrisation with the BK
parametrisation. Both representations have a pole in addition to $D_s^*$, but
the  former fixes it at threshold while the latter keeps this
pole at $t=m_{D_s^*}/\alpha$ as a free parameter. This effective pole sums up 
the contributions of a tower of resonances to the cut. We find that $\alpha<1$, consistently
with the location of the cut. Let us also recall that one should have $\alpha
\to 1$ in the heavy-quark limit and large-energy
release limit: therefore, 
it is consistent to get something rather different from
$1$ in the decay $D\to K\ell\nu$~\footnote{The BK type of parametrisation can
  be used, and has indeed been used prior to the recent experimental
  measurement, in order to represent model or lattice QCD calculations 
of $D \to K \ell \nu$ in the physical region, and deduce
the residue at the pole. The authors of refs.~~\cite{Melikhov:2000yu,Melikhov:2001zv} 
found $\alpha=0.24$ from  a quark model, 
and a lattice QCD computation~\cite{Abada:2000ty} led to
$\alpha=0.27(11)$ or $0.43(12)$, to be compared with the present $0.38$.}.

\section{The strong coupling $g_{D_s^* D K}$} \label{ghat}

Extrapolating the results of the previous section 
to the $D_s^*$ pole and propagating the errors, we obtain $g_{D^*_s DK}$
through eq.~(\ref{eq:poleg}), which is 
related to the matrix element of the axial current between the $B$ and $B^{*}$
mesons. After a suitable normalisation of states, 
this matrix element is denoted as $\hat{g}$ and one has~\footnote{
This relation is based on the heavy-quark expansion, but also on 
PCAC for the kaon and thus could be affected by corrections of order 30\%.}:
\begin{equation}
\hat{g}=\frac{f_K g_{D^*_s DK}}{\sqrt{m_D m_{D_s^*}}}
\end{equation}
$\hat{g}$ is finite in the heavy-quark mass limit, and thus should vary slowly
with the heavy-quark mass.  Another
important property is that the strong couplings of the $K$ meson  (Goldstone
boson) with the $D$ and any other $D_s$-type state satisfy an Adler-Weisberger
sum rule, with all the intermediate states contributing positively,
yielding a bound on the $g_{D_s^* D K}$ coupling. In terms of $\hat{g}$, this
bound is quite simple:
\begin{equation}
|\hat{g}| < 1
\end{equation}
The upper bound cannot be reached because there are indeed transitions from
the $D$ meson to other states than the $D_s^*$. The limit $1$ corresponds to a
completely non-relativistic calculation (for more details on
  $\hat{g}$, see for instance ref.~\cite{Becirevic:1999fr}).

Numerically, we take $f_K=154$ MeV, $m_D=1.864$ GeV,
$m_{D_s^*}=2.112$~GeV. Since 
$f_{D_s^*}$ has not been measured directly, we must rely
on lattice calculations. $N_F=0$ calculations from the Rome
\cite{Becirevic:2000kq} and UKQCD \cite{Bowler:2000xw} lattice collaborations
are available, with sizably different numbers. Unfortunately, neither 
papers quote their value of $f_{D_s^*}$ explicitly. 
In the case of the Rome collaboration, we have to pass through the ratio 
$f_{D_s^*}/f_{D^*}=1.10$ and the value of $f_{D^*}=258(14)(6)$~MeV. 
In the case of UKQCD, we have to convert their result from their dimensionless 
definition of $f_{D_s^*}=8.3$ to our definition
$f_{D_s^*}=m_{D_s^*}/f_{D_s^*}^{UKQCD}$. In addition, it must be noticed that
the mass $m_{D_s^*}$ must be taken consistently from the lattice data at the same
$\beta$ and with the same way of fixing the lattice unit as  $f_{D^*}$. We choose
$\beta=6.2$ and we use $f_{\pi}$ to fix the lattice unit (as done to compute
$f_{D_s^*}$) so that $m_{D_s^*}=2.061$~GeV. Following this procedure, we obtain
finally the central values $f_{D_s^*}=284$ MeV (Rome) and $f_{D_s^*} \simeq
248$~MeV (UKQCD).

A very naive average yields the value $f_{D_s^*}=270$~MeV that we use in the
following. We do not quote errors on the auxiliary quantity $f_{D_s^*}$: it would represent
a very involved task for a quantity that we use only as a reference scale to compare
different parametrisations. However, one should keep in mind that the value of
this decay constant may be underestimated in view of the situation for the closely
related quantity $f_{D_s}$. Indeed the same lattice groups predicted the latter around $230$~MeV,
while the most recent experimental average~\cite{Rosner:2008yu} yields
$275 \pm 10$~MeV~\footnote{The PDG Review of
particle properties 2006  was quoting a still higher value:
$f_{D_s}=294 \pm 27$~MeV.}. If $f_{D_s^*}$ were to be enhanced, $g_{D^*_s DK}$
and $\hat{g}$ should be accordingly rescaled (and lowered). However, the hierarchy observed
in the results of the extrapolations will not be affected by this change in
the overall normalisation.

We have discarded the simple pole model in the discussion
of the residue: it does not account for the data properly, even with a
flexible vector meson mass, and it would be meaningless to
discuss the residue with a fictitious pole mass.

\begin{table}[t]
\begin{center}
\begin{tabular}{ccccc}
Model & No radiative corr & First bin corrected & 
               Full radiative corr\\
\hline
Unitary $z$-exp 
        & $ 16.2\pm 7.8$ & $ 17.9\pm 7.8$ & $ 23.0\pm 8.7$\\
BK      & $ 18.3\pm 1.1$ & $ 18.2\pm 1.0$ & $ 18.2\pm 1.2$\\
Linear  & $ 16.3\pm 0.6$ & $ 16.2\pm 0.6$ & $ 16.1\pm 0.6$\\
Quadratic 
        & $ 16.1\pm 6.5$ & $ 16.6\pm 6.4$ & $18.1\pm 6.4$\\
Simple $z$-exp
        & $ 15.4\pm 3.2$ & $ 16.1\pm 3.2$ & $18.2\pm 3.1$\\
\hline
\end{tabular}
\end{center}
\caption{Values of $g_{D^*_s DK}$ obtained from different parametrisations of $f_+$,
  with various treatments of radiative
corrections (not included, partially included, fully included).}
\label{tab:gDstarDK}
\end{table}

\begin{table}[t]
\begin{center}
\begin{tabular}{ccccc}
Model & No radiative corr & First bin corrected & 
               Full radiative corr\\
\hline
Unitary $z$-exp 
        & $ 0.63\pm 0.31$ & $ 0.70\pm 0.30$ & $ 0.89\pm 0.34$\\
BK      & $ 0.71\pm 0.04$ & $ 0.70\pm 0.04$ & $ 0.71\pm 0.05$\\
Linear  & $ 0.64\pm 0.03$ & $ 0.63\pm 0.03$ & $ 0.62\pm 0.03$\\
Quadratic 
        & $ 0.62\pm 0.25$ & $ 0.64\pm 0.25$ & $0.70\pm 0.25$\\
Simple $z$-exp
        & $ 0.59\pm 0.12$ & $ 0.62\pm 0.13$ & $0.71\pm 0.13$\\
\hline
\end{tabular}
\end{center}
\caption{Values of $\hat{g}$ obtained from different parametrisations of $f_+$,
  with various treatments of radiative
corrections (not included, partially included, fully included).}
\label{tab:ghat}
\end{table}

The results for $g_{D^*_s DK}$ and $\hat{g}$ are collected in
Tables~\ref{tab:gDstarDK} and~\ref{tab:ghat}. Models describing 
the singularities of the cut in an appropriate way should yield
values of $\hat{g}$ in good agreement with what has been obtained
from experimental measurement of strong decays or from the lattice
calculations.

The only indication from experiment is indirect because it concerns the
$SU(3)$ counterpart of $g_{D^*_s DK}$, $g_{D^* D \pi}$. Indeed,
CLEO~\cite{Ahmed:2001xc,Anastassov:2001cw} 
has measured the decay width of $D^* \to D \pi$, from which they estimate:
\begin{equation}
g_{D^* D \pi}=17.9 \pm 0.3 \pm 1.9
\end{equation}
leading to:
\begin{equation}
\hat{g}=0.59 \pm 0.07 \simeq 0.6
\end{equation}
The latter value should apply roughly for the $D^*_s -D$ transition through
$SU(3)$ flavour symmetry (the Dirac quark model and the lattice results suggest an
increase when one increases the light quark mass up to the strange mass). 

Model approaches can be used to determine this coupling. The Dirac model, see for example
ref.~\cite{Becirevic:1999fr}, led to a result close to $\hat{g} \simeq 0.6$ for a
\emph{static} $c$ quark, a number close to what was found later in the above experiment\footnote{A
  careful study based on QCD sum rules and including radiative corrections
  \cite{Khodjamirian:1999hb}
 yields a much lower number. A possible way of
curing this surprisingly small result has been proposed in
\cite{Becirevic:2002vp}:
on the hadronic side, a large contribution from 
radial excitations should be added
to the standard perturbative continuum. Alternative estimations from
light-cone sum rules for semileptonic $B$ decays~\cite{Ball:2004ye} point
towards values of $\hat{g}$ closer to the ones collected here.}. 
The application of a dispersive approach to a constituant quark model led
to $\hat{g}$ from 0.4 to 0.5~\cite{Melikhov:1999nv,Melikhov:2001zv}.

The result of lattice QCD simulations for the $D^* -D$ transition is, 
by extrapolation to the $u,d$ mass (ref. \cite{Abada:2002vj}):
\begin{equation}
\hat{g}=0.67(8)^{+4}_{-6}
\end{equation}
which is compatible with the experimental number quoted above.
The lattice can in fact measure directly another
$SU(3)$ 
partner of the $D^*_s-D$ axial transition matrix element (corresponding to the $A_1$ form factor, up to a mass factor very close to $1$), i.e.
$D^*_s-D_s$, since the lattice direct measurement is close to the strange
quark mass; the result is close to $0.7$ from same reference, at $\beta=6.2$. 

Comparing these values (all in the same range) 
with the above tables, which collect the
coupling obtained by extrapolation of $D \to K \ell \nu$, 
we can make the following comments:
\begin{itemize}
\item
The result from the (unitary) $z$-expansion is sensitive to the extraction 
of the radiative corrections on the whole range of momenta. With a full
treatment of radiative corrections, one obtains a neatly larger number than with the other
parametrisations, and the central value of $\hat{g}$ is close to the
upper bound set by the Adler-Weisberger sum rule. We interpret this as a
consequence of the spurious pole at $t=t_+$: this
representation must be discarded when the unphysical region is concerned
because of its unphysical behaviour at the threshold of the cut.

\item The four other results are rather close to each other which is
  encouraging, since they rely on varied but reasonable assumptions on the 
smoothness of the cut. In particular, the simple $z$-expansion yields a value
in good agreement with the other models, confirming that the spurious pole in
$\Phi$, imposed by unitarity constraints, is the actual source of difficulties
for the unitary $z$-expansion.

\item For these four models,
the central value is in agreement with the independent determinations of
$\hat{g}$, in particular the experimental one. This provides further support
for our preferred assumptions and parametrisations.

\item One can check a posteriori that the effect of the cut is varying
  very slowly with $t$. 
Let us take the linear parametrisation as an illustration.
After subtracting the pole which is at its right position and strength,
the remaining contribution to the form factor is just a constant, that is, \emph{the cut is represented by a
 constant $c$}, with $c_1= -0.42 f_+(0)$ (the negative sign indicates that it
has a lowering effect on the form factor). With such a parametrisation, we are
 sensitive to the structure of the cut neither in the physical region - which is
perhaps not too surprising - nor in a large part of the unphysical
region $t_{-}<t<t_+$. 

\item Comparing the "quadratic" and "linear" fits, we observe that a
  two-parameter fit, which seems equally reasonable as a one-parameter
  parametrisation, yields no change in the central value of $\hat{g}$; 
    however, the errors are
  increased leading to a loss of predictive power. This means
  that the parametrisations are not constrained enough by the data in the
  physical region for a compelling extrapolation.
\end{itemize}

\section{Conclusions}

We have exploited recent high-precision data of the BaBar collaboration on
the $D\to K\ell\nu$ decay in order to test various parametrisations of
heavy-to-light form factors. Indeed, the current theoretical description of
such form factors remains largely incomplete, and it can be improved by a
direct comparison with data. The accuracy of the experimental results yields 
good fits of 
these parametrisations of the vector form factor $f_+$ in the physical
region. But the value of the vector form factor outside this region is also of
interest, since its residue at the mass of the $D_s^*$ meson is related to the
strong coupling constant $\hat{g}$. To extract this quantity, we need to
extrapolate the form factors outside the semileptonic region, and thus to rely
on the properties and assumptions of the various parametrisations for $f_+$.
On the other hand, the determination of $hat{g}$ and its comparison with
values from other approaches (measurements of strong decays, lattice
computations) provides an interesting test of the parametrisations of the form
factors, since it probes the differences in the way these parametrisations 
representation the physical singularities along the cut.

In spite of the modesty of our approach, which is very phenomenological, some useful conclusions can be drawn: 
\begin{itemize}
\item The method of extrapolation using \emph{smooth} models for the cut -
  either a \emph{remote} effective pole (BK) or a low-order polynomial in $t$ -
  has an encouraging success, since the $D_s^*$ residue appears quite
  compatible with values expected from very different considerations. The $z$
  expansion (in its unitary version, widely popular by now) is disfavored:
  it contains a spurious pole at the threshold of the $DK$ cut which makes the
  extrapolation blow out of control in the unphysical region.

\item The uncertainty on $f_{D_s^*}$ is still large; reducing it would in
  turn reduce that of the strong coupling constant and help to strengthen the
  previous conclusion.

\item The extrapolation has by itself a very large uncertainty, which cannot
  be reduced further if we know only the physical region. Some quantitative
  theoretical knowledge about the form factor on the cut , in particular at
  threshold (scattering length\ldots), 
  would be a great help by transforming the extrapolation into an
  interpolation. Borrowing ideas from ref.~\cite{Flynn:2007ki}, one could then
  use a sufficiently subtracted Omn\`es-Muskhelishvili representation to
  combine our knowledge on the cut near $DK$ threshold and the physical
  semileptonic region.

\item Another improvement would consist in enlarging the interval of the
  fit. Such information could be provided by lattice simulations computing the form
  factor for $t<0$ (indeed they can calculate the $D-K$ matrix element without 
  referring to the $D \to K \ell \nu $ transition). However, the improvement
  would be much milder, since the sensitivity to the cut decreases when one
  gets deeper into the region of negative $t$.

\end{itemize}

The $D \to K \ell \nu $ process is a particular decay among many similar semileptonic
processes. It presents the advantage of a highly accurate knowledge of the
form factor, and of a clear analytical structure, with the physical threshold,
the pole and the cut neatly separated, which will not be the case for other
processes. Our simple analysis of this particular process shows some
interesting features of the different parametrisations of the form factors
currently used to analyse weak transitions from one meson to another. In
particular, it should invite practitioners 
to proceed with care when they have to rely heavily on 
parametrisations to extract quantities of physical interest
(such as CKM matrix elements or strong hadronic couplings) 
with a high accuracy.

\vspace*{1em}

\noindent
\subsubsection*{Acknowledgments}

We thank Patrick Roudeau (LAL) who suggested and initiated this study, and who provided
us many precious indications on the BaBar measurements of the form factor, as
well as Damir Be\'cirevi\'c (LPT) for his numerous comments and discussions. Our work has
also benefited from the various joint discussions between LAL experimentalists and LPT theorists.

This work was supported in part by the EU Contract No.
 MRTN-CT-2006-035482, \lq\lq FLAVIAnet'' and by the ANR project QCDNEXT (ANR\_NT05-3\_43577).

\bibliographystyle{JHEP}
\bibliography{biblio}

\end{document}